# Decoded fMRI neurofeedback can induce bidirectional behavioral changes within single participants

Aurelio Cortese[1,2,3,4,7], Kaoru Amano[3,7], Ai Koizumi[1,3,4,5,7], Hakwan Lau[4,6] & Mitsuo Kawato[1,2]

1. Dep. of Decoded Neurofeedback, ATR Cognitive Mechanisms Laboratories, Kyoto, Japan
2. Faculty of Information Science, Nara Institute of Science and Technology, Nara, Japan
3. Center for Information and Neural Networks (CiNet), NICT, Osaka, Japan
4. Dep. of Psychology, UCLA, Los Angeles, USA
5. Dep. of Psychology, Columbia University, New York, USA
6. Brain Research Institute, UCLA, Los Angeles, USA
7. These authors contributed equally to this work

correspondence should be addressed to Aurelio Cortese (cortese.aurelio@gmail.com), Hakwan Lau (hakwan@gmail.com), or Mitsuo Kawato (kawato@atr.jp)

# Abstract

Studies using real-time functional magnetic resonance imaging (rt-fMRI) have recently incorporated the decoding approach, allowing for fMRI to be used as a tool for manipulation of fine-grained neural activity. Because of the tremendous potential for clinical applications, certain questions regarding decoded neurofeedback (DecNef) must be addressed. Neurofeedback effects can last for months, but the short- to mid-term dynamics are not known. Specifically, can the same subjects learn to induce neural patterns in two opposite directions in different sessions? This leads to a further question, whether learning to reverse a neural pattern may be less effective after training to induce it in a previous session. Here we employed a within-subjects' design, with subjects undergoing DecNef training sequentially in opposite directions (up or down regulation of confidence judgements in a perceptual task), with the order counterbalanced across subjects. Behavioral results indicated that the manipulation was strongly influenced by the order and direction of neurofeedback. We therefore applied nonlinear mathematical modeling to parametrize four main consequences of DecNef: main effect of change in behavior, strength of down-regulation effect relative to up-regulation, maintenance of learning over sessions, and anterograde learning interference. Modeling results revealed that DecNef successfully induced bidirectional behavioral changes in different sessions. Furthermore, up-regulation was more sizable, and the effect was largely preserved even after an interval of one-week. Lastly, the second week effect was diminished as compared to the first week effect, indicating strong anterograde learning interference. These results suggest reinforcement learning characteristics of DecNef, and provide important constraints on its application to basic neuroscience, occupational and sports trainings, and therapies.

# Introduction

Real-time functional magnetic resonance imaging (rt-fMRI) neurofeedback has enjoyed a considerable rise in interests in recent years, due to its great potential in investigating scientific questions via direct manipulation of specific brain activity patterns, as well as potential clinical applications (Sulzer et al. 2013; deCharms 2008). Whereas most previous studies have mainly focused on participants learning to self-regulate a univariate blood-oxygen-dependent-level (BOLD) signal in specific brain areas (Weiskopf et al. 2003; deCharms et al. 2004; Birbaumer et al. 2013; Sulzer et al. 2013), only recently has multi-voxel pattern analysis (MVPA) or decoding analysis (Kamitani & Tong 2005) been applied to rt-fMRI neurofeedback, opening a new range of possibilities (Shibata et al. 2011; deBettencourt et al. 2015).

Previously, up-and-down regulation of univariate BOLD signal within a single subject in interleaved block designs of rt-fMRI neurofeedback has been shown to be possible under several conditions (Weiskopf et al. 2004; deCharms 2008). These conditions include regulation of BOLD signals within specific ROIs and differences in activation between two ROIs; furthermore, in most rt-fMRI neurofeedback studies, participants were given specific explicit strategies of neural induction as well as neural loci through verbal instructions. For example, Scheinost *et al.* showed that up-and-down regulation of orbitofrontal cortex (OFC) is beneficial for OCD therapy, although they did not examine each effect separately for activation and deactivation (Scheinost et al. 2013). Similarly, in a different study, subjects learned to voluntarily up- and down-regulate the activity level of the anterior insula (Veit et al. 2012). The aim was analyzing the functional interactions between different brain areas during rt-fMRI neurofeedback, but neither study examined the effect of up or down-regulation separately on functional connectivity or behaviors. In contrast, Shibata and colleagues changed cingulate cortex multi-voxel patterns by decoded neurofeedback (DecNef) in two opposite directions for two different groups of participants and observed increase and decrease of facial preference in each group, respectively (Shibata et al. 2016). However, these differential behavioral results caused by differential multi-voxel manipulations were observed for different groups of participants, but not within single participants.

From a broader perspective, considering traditional ROI-based rt-fMRI neurofeedback, connectivity neurofeedback (FCNef) (Megumi et al. 2015), and DecNef, to the best of our knowledge, no study has successfully demonstrated opposing behavioral outcomes for different neurofeedback manipulations within single subjects. This open question is important for both practical as well as basic neurobiological reasons. Neurofeedback is a powerful research tool for human systems neuroscience because it can address causal relationships between altered brain dynamics and resulting behaviors. As in optogenetics studies in rodents, to rigorously prove the causal relationships between brain activity and behaviors, it is desirable to be able to induce and suppress the same pattern of brain activities, and confirm that they indeed lead to opposite behavioral effects within the same subject. Therefore, studying the extent and dynamics of DecNef's effects, as well as their mechanisms, is of central importance for the development of the technique for clinical and rehabilitative applications.

We aimed to address the following three questions in this research. First, if some behavioral change is induced by a neurofeedback manipulation, is it possible to develop another neurofeedback manipulation to cancel out the first behavioral change? Second, how long is the behavioral change maintained after neurofeedback manipulation? Third, how much interference occurs when two different neurofeedback manipulations are conducted in single participants? The first question is important, as it is desirable to have the ability to cancel out presumed negative side effects in the event these occur. The second question is related to efficiency of neurofeedback as a therapeutic method. Megumi *et al.* (Megumi et al. 2015) showed that 4 days of FCNef changed resting-state functional connectivity, and these lasted more than two months. Amano *et al.* (Amano et al. 2016) showed that 3 days of DecNef induced associative learning between color and orientation lasts for 3 to 5 months. However, there is no quantitative study examining mid-term effects; for example, to what extent are neurofeedback effects maintained one week after manipulation? The third point is ethically important in considering cross-over designs as candidate paradigms for randomized control trials to show statistical effectiveness of neurofeedback therapy. If two different neurofeedback manipulations interfere severely within single patients, a cross-over design is not a feasible option.

DecNef and FCNef can be assumed as a neural operant conditioning or reinforcement learning process, since it is essentially a reward-based manipulation of brain states and participants have no conscious understanding of the objectives of neurofeedback (Bray et al. 2007; Shibata et al. 2011; Megumi et al. 2015). Furthermore, unlike most of rt-fMRI neurofeedback studies, DecNef and FCNef do not utilize verbal instructions about conscious strategy to participants, and thus likely depend on more automatic and implicit processes compared with other traditional neurofeedback paradigms. Generally speaking for fMRI neurofeedback, basal ganglia are thought to be amongst the key areas playing a pivotal role in the procedural nature of neurofeedback (Birbaumer et al. 2013; Koralek et al. 2012). Since DecNef relates to learning, an interesting argument can be explored: interference of learning is expected when we attempt a two-way behavioral manipulation. Disruption of learning has been studied extensively in motor and visuomotor learning, with various elegant studies showing that previous learning can hinder or interfere with the subsequent practice of a second task (Brashers-Krug et al. 1996; Krakauer et al. 1999; Osu et al. 2004). Interference can be retrograde or anterograde, depending on the direction in time of the learning/memory effects. In a classic $A_1BA_2$ paradigm, subjects are instructed to sequentially learn Task A, Task B, and then Task A again. Retrograde effects reflect how learning of Task B affects the memory of task A, while anterograde effects reflect how the memory of task A affects the learning of task B (Sing & Smith 2010). Anterograde interference has received less attention in the literature (Sing & Smith 2010). Interference effects have also been shown to be present in perceptual learning (Seitz et al. 2005; Yotsumoto et al. 2009). Therefore, anterograde interference of learning resulting from the bidirectional use of DecNef manipulations within subjects may prove two aspects: (1) behavioral changes are the result of a true learning process and, (2) behavioral effects should be long lasting.

Here we used DecNef to manipulate a specific cognitive property - perceptual confidence - bidirectionally, i.e., up (increase confidence), and down (decrease confidence). We first constructed a classifier for high versus low confidence by utilizing MVPA. In a second stage, with a within-subjects design, participants learned to implicitly induce multi-voxel activation patterns reflecting high and low confidence levels over two weeks. In both weeks, induction sessions took place across two consecutive days, and behavioral changes were measured with a Pre- and Post-Test, immediately before and after induction session. Participants were randomly assigned to one of two groups, defining the order of induction (Up- then Down-DecNef or vice versa). The second DecNef session was carried one week after the first, in order to measure whether effects would survive a one week interval. To best capture the differential effects of DecNef on confidence judgements, we utilized nonlinear equation modeling.

# Materials and methods

## Subjects and Experiment Design

All experiments and data analyses were conducted at the Advanced Telecommunications Research Institute International (ATR). The study was approved by the Institutional Review Board of ATR. All subjects gave written informed consent. A companion paper (Cortese *et al.*, submitted) was submitted elsewhere and discussed implications of the results for neural mechanisms of meta-cognitive function, especially perceptual confidence. The experimental data used is exactly the same, but the research objectives of the two manuscripts are not overlapping.

The cognitive aspect that served as the basis for the computational analysis of DecNef effect was perceptual confidence. As a working definition of confidence, we can intend it as the degree of certainty in one's own perceptual decisions. Perceptual confidence can take graded levels between a low (uncertain), and high (certain) states. For experimental design purposes, data samples were subdivided between the two confidence boundaries, thus creating two categories: Low and High Confidence.

The entire experiment was subdivided into 6 fMRI sessions (Fig. 1A). In the first session, participants did a retinotopy scan to functionally define visual areas, followed by an MVPA session, on separate days. After successful MVPA, where decoding of confidence attained more than 55% cross-validated accuracy in any of the frontoparietal ROIs of interest (see below), participants were randomly assigned to one of two groups, with respect to the order of the confidence inductions in DecNef: Down-Up group (aiming at Low- then High confidence, D-U throughout the manuscript) or Up-Down group (aiming at High- then Low confidence, U-D throughout the manuscript). The behavioral performance in these tasks, before vs after DecNef, is our primary dependent variable of interest. Eighteen subjects (23.7±2.5 years old; 4 females and 14 males) with normal or corrected-to-normal vision participated in the first part of the study (retinotopy mapping and MVPA). Ten subjects with the highest confidence decoding accuracy in frontoparietal ROIs were selected to perform the full DecNef experiments (24.2±3.2 years old, 3 females and 7 males). The screening was opted for based on previous DecNef experiments (to ensure successful neurofeedback induction, decoding accuracy needs to be higher than ~60%). Furthermore, DecNef being a causal tool, if the independent variable is manipulated by the experimenter (in this case, decoding accuracy higher than a certain threshold), if it is truly representing the encoding of confidence, a change in the dependent variable in Pre- and Post-tests will be seen.

Stimuli, Behavioral and DecNef Designs

The behavioral task was a two-alternative forced choice of motion discrimination with confidence rating using random dot motion (RDM) (Fig. 1B). Subjects were instructed to indicate the direction of motion (left or right) after a short delay following stimulus presentation, and rate the level of confidence on their decision (4-point scale). Performance in judging the motion direction was not significantly different from the target level of 75% correct (76.8%±1.6%). A binary confidence decoder for High vs Low confidence was created by pooling together samples from the lower half and higher half of confidence choices distribution, individually. Specifically, we re-assigned the intermediate levels (2, 3) to both the low and high confidence classes in order to collapse the 4 confidence levels into 2 levels, and equate the number of trials in each class. First, one intermediate class was merged with the high or low confidence group, depending on the total number of trials. To equate the number of trials, randomly sampled trials from the left-out intermediate confidence class were added to the confidence group now having a lower total number of trials. The balancing was based on the confidence rating response distribution, and on the final number of trials. To construct the binary classifiers, we used sparse logistic regression (SLR), which automatically selects relevant voxels for classification (Yamashita et al. 2008). We constructed four decoders in frontoparietal areas, corresponding to the following anatomical ROIs: inferior parietal lobule (IPL), and three subregions generally regarded as being part of the dorsolateral prefrontal cortex, namely the inferior frontal sulcus (IFS), middle frontal sulcus (MFS), and the middle frontal gyrus (MFG). These areas have been previously linked to confidence judgements in perceptual decisions (Kiani & Shadlen 2009; Fleming et al. 2010; Rounis et al. 2010).

Once individual confidence classifiers were constructed, each participant completed a two-day DecNef training in each session (Down/Up), and each session was separated by at least one week (see Fig. 1A). Each subject went through both DecNef training for High and Low confidence, and the order of confidence inductions was counterbalanced across subjects (i.e., Up then Down vs Down then Up). The neurofeedback task itself is illustrated in Fig. 1C: participants were asked to "manipulate, modulate or change their brain activity in order to make the feedback disc presented at the end of each trial as large as possible". The experimenters provided no further instructions nor strategies. Without knowing the actual meaning of the feedback, subjects learned to implicitly induce brain activation patterns corresponding to high or low confidence. Participants received monetary reward proportional to their induction success (ability to implicitly induce the selected activation pattern).

After each scanning day, subjects were asked to describe their strategies in making the disc size larger. Answers varied from "I was counting", to "I was focusing on the disc itself", to "I was thinking about food". All answers can be found in appendix A. When subjects were asked about which group they thought they were assigned to at the end of the experiments (N=5, 2 months later, and N=4, 5 months later - 1 subject could not be joined), their answers were at chance (57% correct, Chi-square test, $\chi^2 = 0.225$, $P = 0.64$).

The details of the DecNef methodology have been published previously (Shibata et al. 2011). On each day of a given DecNef session, subjects participated in up to 11 fMRI runs. The mean (±s.e.m) number of runs per day was 10±0.1 across days and subjects. Each fMRI run consisted of 16 trials (1 trial=20 sec) preceded by a 30-sec fixation period (1 run=350 sec). The fMRI data for the initial 10 sec were discarded to avoid unsaturated T1 effects.

Each trial started with a visual cue (three concentric disks, white, gray and green) signaling the induction period. Induction lasted 6 sec, and was followed by a 6 sec rest period, before the neurofeedback disk (a white ring) was presented on the gray screen for up to 2 sec. Finally, a trial ended with a 6 sec ITI.

During the fixation period, subjects were asked to simply fixate on the central point and rest. This period was inserted between the induction and the feedback periods to account for the hemodynamic delay, assumed to last 6 sec. The following feedback period corresponded to a 2 sec presentation of a white disc. The size of the disc represented how much the BOLD signal patterns obtained from the induction period corresponded to activation patterns of the target confidence state (high or low). The white disc was always enclosed in a larger white concentric circle (5 deg radius), which indicated the disc's maximum possible size.

The size of the disc presented during the feedback period was computed at the end of the fixation period according to the following steps. First, measured functional images during the induction period underwent 3D motion correction using Turbo BrainVoyager (Brain Innovation) for each of the four frontoparietal ROIs used (IPL, IFS, MFS, and MFG). Second, time-courses of BOLD signal intensities were extracted from each of the voxels identified in the MVPA session, and were shifted by 6 sec to account for the hemodynamic delay. Third, a linear trend was removed from the time-course, and the BOLD signal time-course was z-score normalized for each voxel using BOLD signal intensities measured for 20 sec starting from 10 sec after the onset of each fMRI run. Fourth, the data sample to calculate the size of the disc was created by averaging the BOLD signal intensities of each voxel for 6 sec in the induction period. Finally, the likelihood of each confidence state was calculated from the data sample using the confidence decoder computed in the MVPA session. The size of the disc was proportional to the averaged likelihood from the four different frontoparietal ROIs (ranging from 0 to 100%) of the target confidence assigned to each subject on a given DecNef block. Importantly, subjects were unaware of the relationship between their activation patterns induction and the size of the disk itself. The target confidence was the same throughout a DecNef block. In addition to a fixed compensation for participation in the experiment, a bonus of up to 3000 JPY was paid to the subjects based on the mean size of the disc on each day.

All stimuli were created and presented with Matlab (Mathworks) using the Psychophysics Toolbox extensions Psychtoolbox 3 (Brainard 1997). The behavioral task was the same for both MVPA, and Pre- Post-Tests. In behavioral studies, subjects performed the task on a standard computer and gave responses from a standard keyboard. In the fMRI sessions, subjects gave their responses via a 4-buttons pad.

Visual stimuli were presented on an LCD display (1024 × 768 resolution, 60Hz refresh rate) during titration and the Pre- and Post-Test stages, and via an LCD projector (800 × 600 resolution, 60Hz refresh rate) during fMRI measurements in a dim room. Stimuli were shown on a black background and consisted of RDM. We used the Movshon-Newsome (MN) RDM algorithm (Shadlen & Newsome 2001). The stimulus was created in a square region of $20 \times 20$ deg, but only the region within a circular annulus was visible (outer radius: 10 deg, inner radius: 0.85 deg). Dot density was 0.5 deg$^{-2}$ (contrast 100%), with a speed of 9 deg/sec and size of 0.12 deg. Signal dots all moved in the same direction (left or right, non-cardinal directions of 20 deg and 200 deg) whereas noise dots were randomly replotted. Dots leaving the square region were replaced with a dot along one of the edges opposite to the direction of motion, and dots leaving the annulus were faded out to minimize edge effects.

## Mathematical modeling and model comparison

We constructed a mathematical model to objectively examine the effects of Up and Down DecNef, a decay of learning (of DecNef effect) due to one-weak elapse, and an anterograde interference of learning from the first to second weeks. The model was fit to 4 measurement values of perceptual confidence at the 4 time points for each participant and possesses 4 model parameters. $X_j^i$ are the experimentally measured confidence values, while $\widehat{X}_j^i$ are the estimated confidence values, for each subject (with $i = 1:4$) (i.e., behavioral outcome of DecNef effects) at each time point (with $j = 1:4$); The main 4-parameter nonlinear model to describe DecNef effects is formally outlined as follows.

$$\widehat{X}_1^i = B \tag{1}$$

*for* $i = 1:5,$ group D-U

$$\widehat{X}_2^i = B + \varepsilon \cdot \Delta \tag{2}$$

$$\widehat{X}_3^i = B + \varepsilon \cdot \Delta \cdot \alpha \tag{3}$$

$$\widehat{X}_4^i = B + \varepsilon \cdot \Delta \cdot \alpha + \gamma \cdot \Delta \tag{4}$$

*for* $i = 6:10,$ group U-D

$$\widehat{X}_2^i = B + \Delta \tag{5}$$

$$\widehat{X}_3^i = B + \Delta \cdot \alpha \tag{6}$$

$$\widehat{X}_4^i = B + \Delta \cdot \alpha + \varepsilon \cdot \gamma \cdot \Delta \tag{7}$$

$$Error = \Sigma^i \Sigma^j \left( X_j^i - \widehat{X}_j^i \right)^2 = F(\alpha, \varepsilon, \gamma, \Delta) \tag{8}$$

Where $B$ is the initial baseline (0 - the realigned confidence level); Δ, the change in confidence by Up-DecNef in the first week; ε, the ratio of Down-DecNef effect normalized by Up-DecNef effect; ɣ, the anterograde learning interference resulting in reduced second week DecNef effect; and ɑ, the learning persistence ([1- (decay of learning)] in the week-long interval) during one week. Submodels are defined by setting different parameters to zero or one, one at a time or concomitantly following a complexity logic. This gives rise to a hierarchical group of models, from simpler to most complex (capturing single or increasingly more aspects of DecNef effects on confidence). The first model is the simplest and only estimates Δ, with the other parameters setting as ε = 0, ɣ = 1, ɑ = 1. The second model, by complexity order, assumes Up and Down-DecNef effects, estimates Δ and ε, while ɣ = 1, ɑ = 1. The third model estimates Δ, ε and ɣ, with ɑ = 1. Further DecNef-based models estimate Δ, ε and ɑ, with ɣ = 1; or estimate Δ, ε and ɑ, with ɣ = 0; or estimate Δ, ε and ɣ, with ɑ = 0; or finally, estimate Δ, ɣ, and ɑ, with ε = 0.

We considered alternative models that do not take into account DecNef direction assumptions or a posteriori conceptions. These are a 1-parameter constant confidence model (confidence does not change, is constant throughout the experiment), with two versions: $k = \overline{X}_1^i$, or $k = mean(\overline{X}_2^i, \overline{X}_3^i, \overline{X}_4^i)$. Other free models are a *within-week* constant confidence, and *within-week* constant confidence with two or four additional linear parameters.

For model comparison, we used the Akaike Information Criterion (AIC) (Akaike 1974). Raw AIC is computed according to the following equation:

$$AIC = nlog(\hat{\sigma}^2) + 2k \tag{9}$$

where $\hat{\sigma}^2 = \frac{Residual\ Sum\ of\ Squares}{n}$, $n$ is the sample size and $k$ the number of parameters in the model. In our set of global models, n=30, and k varied from 1 to 4. In the modeling reported for small sample sizes (i.e., n/k <~ 40), the second-order or corrected Akaike Information Criterion (AICc) should be used instead. Although the AICc formula assumes a fixed-effects linear model with normal errors and constant residual variances, while our models are nonlinear, the standard AICc formulation is recommended unless a more exact small-sample correction to AIC is known (Burnham & Anderson 2002):

$$AICc = AIC + \frac{2 \cdot k \cdot (k+1)}{(n-k-1)} \tag{10}$$

For comparing models, two useful metrics are $\Delta_{AICc}$ and Akaike weights ($w_i$). $\Delta^i_{AICc}$ is a measure of the distance of each model relative to the best model (the model with the most negative, or lowest, AIC value), and is calculated as:

$$\Delta^i_{AICc} = AICc_i - min(AICc) \tag{11}$$

As indicated in Burnham and Anderson (Burnham & Anderson 2002), $\Delta^i_{AICc} < 2$ suggests substantial evidence for the *i*th model, while $\Delta^i_{AICc} > 10$ indicates that the model is very unlikely (implausible).

Akaike weights ($w_i$) provide a second measure of the strength of evidence for each model, is directly related to $\Delta^i_{AICc}$ and is computed as:

$$w_i = \frac{exp(-\Delta^i_{AICc}/2)}{\sum_{i=1}^{R} exp(-\Delta^i_{AICc}/2)} \tag{12}$$

AIC analysis results are reported in Table 1, with values reported being AICc, $\Delta_{AICc}$, and Akaike weights ($w_i$).

In cases such as ours, where high degrees of model selection uncertainty exists (the best AIC model is not strongly weighted), a formal solution is to compute parameter estimates through model-averaging. For this approach, two procedures may be used, depending on the results. The first approach makes use of only a limited subset of models that are closest to the current best model ($\Delta_{AICc} < 2$), while the second approach will consider all models (in fact, this accounts to consider all models with $w_i \neq 0$ ). We applied the first approach, selecting only models with high likelihood. Parameters are estimated according to the equation:

$$\widehat{\overline{\beta}} = \frac{\sum_{i=1}^{R} w_i \widehat{\beta}_i}{\sum_{i=1}^{R} w_i} \tag{13}$$

Where $\widehat{\beta}_i$ is the estimate for the predictor in a given model *i*, and $w_i$ is the Akaike weight of that model.

Unconditional error, necessary to compute the unconditional confidence interval for a model-averaged estimate, can be calculated according to the following equation:

$$\widehat{se}(\widehat{\overline{\beta}}) = \sum_{i=1}^{R} w_i \sqrt{\widehat{var}(\widehat{\beta}_i) + (\widehat{\beta}_i - \widehat{\overline{\beta}})^2} \tag{14}$$

Where $\widehat{var}(\widehat{\beta}_i)$ is the variance of the parameter estimate in model *i*, and $\widehat{\beta}_i$ and $\widehat{\overline{\beta}}$ are as defined above. The confidence interval is then simply given by the end points:

$$\widehat{\overline{\beta}} \pm z_{1-\alpha/2} \widehat{se}(\widehat{\overline{\beta}}) \tag{15}$$

For a 90% confidence interval, $z_{1-\alpha/2}$ = 1.65.

Note that the unconditional variance comprises two terms, the first one local (internal variance of model *i*), while the second one global, in that it represents the variance between the common estimated parameter and the true value in model *i*. Since these models are very stable and robust, changing the initial condition set does not lead to different solutions, thus providing proof that solutions reached are globally best. Therefore, in order to assess the variance, we recreated surrogate data sets by selecting $k$ samples out of a population of $n$ samples, with all possible combinations. This is equivalent to the binomial coefficient, thus creating N!/K!(N-K)! subgroups. For each group, parameters were independently estimated, and we thus calculated the population variance.

In the second part of the modeling approach, we consider all data points, and model population's individual fits with nonlinear equations with *global* and *local* parameters. The equations determining the model are thus the same as above :

$$\widehat{X}_1^i = B_i \tag{16}$$

*for i* = 1 : 5, group D-U

$$\widehat{X}_2^i = B_i + \varepsilon \cdot \Delta_i \tag{17}$$

$$\widehat{X}_3^i = B_i + \varepsilon \cdot \Delta_i \cdot \alpha \tag{18}$$

$$\widehat{X}_4^i = B_i + \varepsilon \cdot \Delta_i \cdot \alpha + \gamma \cdot \Delta_i \tag{19}$$

*for i = 6 : 10,* group U-D

$$\widehat{X}_2^i = B_i + \Delta_i \tag{20}$$

$$\widehat{X}_3^i = B_i + \Delta_i \cdot \alpha \tag{21}$$

$$\widehat{X}_4^i = B_i + \Delta_i \cdot \alpha + \varepsilon \cdot \gamma \cdot \Delta_i \tag{22}$$

$$Error = \Sigma^i \Sigma^j \left( X_j^i - \widehat{X}_j^i \right)^2 = F(\alpha, \varepsilon, \gamma, \Delta_i, B_i) \tag{23}$$

Compared with the global-parameter model (Eq. 1-8), in this individualized model $B$ (the initial confidence point) is now optimized individually, as well as $\Delta$, the change in confidence induced by DecNef. For this model, n=40, and k=23.

Analysis, statistics and model-solving routines

All analysis were performed with Matlab (Mathworks) versions 2011b and 2014a with custom made scripts. Additional statistical analysis such as ANOVA were performed with SPSS 22 (IBM statistics). We employed Matlab optimization routines to solve our systems of nonlinear equations with a nonlinear programming solver, under least-square minimization. The Matlab solver was *fmincon*, with the following optimization options. A sequential quadratic problem (SQP) method was used; specifically, the 'SQP' algorithm. This algorithm is a medium-scale method, which internally creates full matrices and uses dense linear algebra, thus allowing additional constraint types and better performance for the nonlinear problems outlined in the previous section. As compared with the default *fmincon* 'interior-point' algorithm, the 'SQP' algorithm also has the advantage of taking every iterative step in the region constrained by bounds, which are not strict (a step can exist exactly on a boundary). Furthermore, the 'SQP' algorithm can attempt to take steps that fail, in which case it will take a smaller step in the next iteration, allowing greater flexibility. We set bounded constraints to allow only certain values in the parameter space to be taken by the estimates, reflecting the biological dimension they were explaining. As such, boundaries were set as: $\Delta \in [0\ \mathrm{Inf}]$, $\varepsilon \in [-1\ 0]$, $\gamma \in [0\ 1]$, and $\alpha \in [0\ 1]$.

The function tolerance was set at $10^{-20}$, the maximum number of iterations at $10^{6}$ and the maximum number of function evaluations at $10^{5}$.

Statistical results involving multiple comparisons are reported in both the corrected and uncorrected forms. The rationale behind this decision is that these comparisons can be interpreted as multiple comparisons of one hypothesis across different mediums, or simply as different hypothesis, in which case no multiple comparisons should be considered. For multiple comparisons, we used the Holm-Bonferroni procedure, where the *P*-values of interest are ranked from the smallest to the largest, and the significance level α is sequentially adjusted based on the formula $\frac{\alpha}{(n-i+1)}$ for the ith smallest *P*-values. Thus, for the analyses with 8 ROIs, the significance levels corresponding to an α of 0.05 were, in increasing order, $α_1$=0.0063, $α_2$=0.0071, $α_3$=0.0083, $α_4$=0.0100, $α_5$=0.0125, $α_6$=0.0167, $α_7$=0.025, $α_8$=0.05. In the text, for enhanced clarity, we present the results as corrected *P*-values.

**MRI parameters**

The subjects were scanned in a 3T MR scanner (Siemens, Trio) with a head coil in the ATR Brain Activation Imaging Center. Functional MR images for retinotopy, the MVPA session, and DecNef stages were acquired using gradient EPI sequences for measurement of BOLD signals. In all fMRI experiments, 33 contiguous slices (TR = 2 sec, TE = 26 ms, flip angle = 80 deg, voxel size = 3×3×3.5 mm$^3$, 0 mm slice gap) oriented parallel to the AC-PC plane were acquired, covering the entire brain. For an inflated format of the cortex used for retinotopic mapping and an automated parcellation method (Freesurfer), T1-weighted MR images (MP-RAGE; 256 slices, TR = 2 s, TE = 26 ms, flip angle = 80 deg, voxel size = 1×1×1 mm$^3$, 0 mm slice gap) were also acquired during the fMRI scans for the MVPA.

# Results

DecNef can be essentially assumed as a neural operant conditioning and/or reinforcement learning paradigm, with which a multi-voxel pattern corresponding to a specific piece of brain information can be induced without explicit knowledge of participants. Throughout the manuscript we refer to Up-DecNef for High-Confidence DecNef and Down-DecNef for Low-Confidence DecNef. Since there were two groups (for DecNef order counterbalancing), we will often refer to these as D-U (first session is Low-Confidence DecNef while the second is High-Confidence) and U-D (the reverse, first High-Confidence DecNef, then Low-Confidence) throughout the results section.

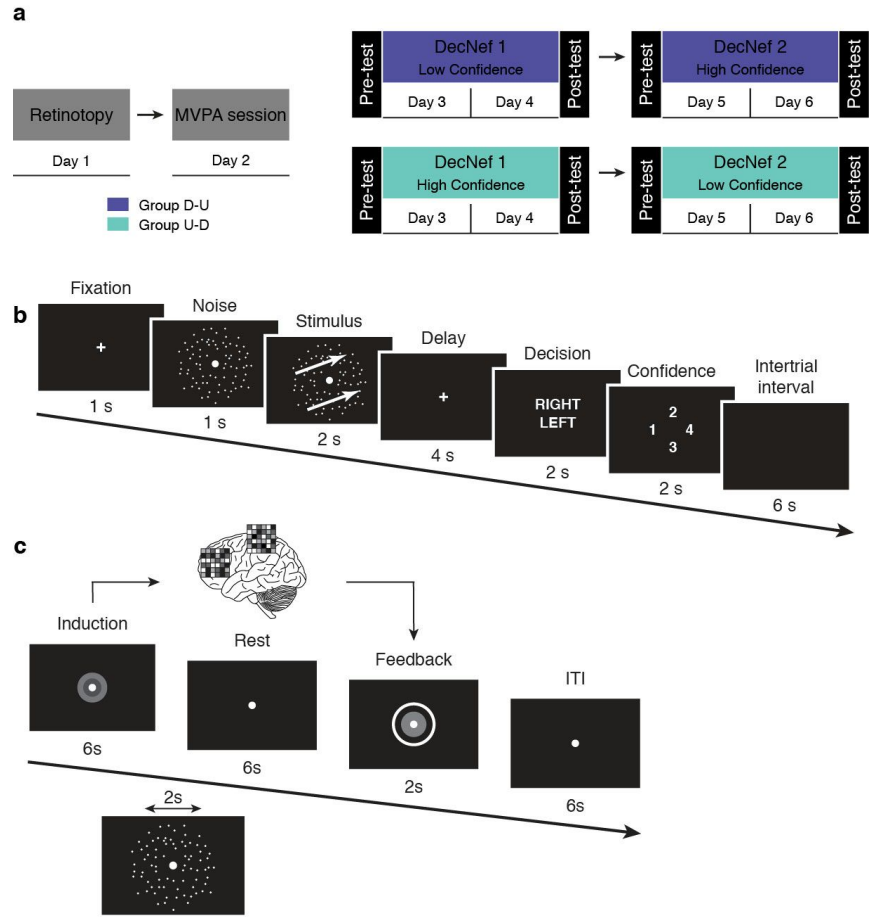

**Figure 1.** Experiment timeline and design. (a) The entire experiment consisted of four sessions divided into 6 fMRI scanning days. The first two days were equal for all groups: a retinotopy session to functionally define visual areas, followed by an MVPA session, during which participants of both groups performed in a 2-forced choice discrimination task with confidence rating. Participants were randomly assigned to either group D-U or group U-D. For the subsequent neurofeedback sessions (day 3-4 and 5-6), group U-D did first High Confidence

followed by Low Confidence DecNef, while group D-U did the reverse sequence. Each DecNef session was preceded and followed by a Pre- and Post-test (on the same days), a psychophysical assessment using the same behavioral task employed in the MVPA session, in order to capture modulations in confidence and perceptual accuracy. (b) In the MVPA session, and in Pre- and Post-Tests, each trial started with a fixation cross, followed by a noise RDM. The stimulus was then presented, consisting of a coherent RDM with either rightward or leftward motion. After a 4 sec delay, subjects were required to give the direction of motion (left or right) and their confidence in their decision during a fixed time window. A trial ended with a 6 sec ITI. 3 TRs, starting at stimulus presentation onset, were averaged and used for the actual MVPA. (c) A neurofeedback trial commenced with a visual cue indicating the induction period, during which subjects were asked to "manipulate, change their brain activity in order to maximize the size of the feedback disc and the reward". Induction was followed by a rest period, then the feedback disc was presented for 2 sec and a trial ended with a 6 sec ITI. Either during the induction period, or at the beginning of the fixation period (pseudo-random onsets: 2, 4, 6, or 8 sec from trial start) a 2 sec noise RDM was also presented. Pseudo-random onsets were designed in order to ensure minimal interference and maximal effect of the RDM on the induction process. Group D-U: Down- then Up-DecNef, group U-D: Up- then Down-DecNef.

---

Behavioral data from the Pre- and Post-Tests show that confidence was differentially manipulated by DecNef (Fig. 2). Importantly, the resulting changes in confidence cannot be attributed to a simple week order effect (ANOVA with repeated measures, non-significant effects of neurofeedback, $F_{1,9}$ = 0.370, $P$ = 0.558, and time, $F_{1,9}$ = 2.834, $P$ = 0.127, and non-significant interaction, $F_{1,9}$ = 0.844, $P$ = 0.382).

As displayed in Fig. 2A, the confidence change is larger for Up-DecNef than Down-DecNef, but importantly, the confidence level attained at the end of the first week was almost entirely preserved until the beginning of the second week, in the second session. Lastly, the second week effect seems present but reduced as compared to the first week effect. Thus, order of DecNef (Up then Down, or Down then Up) had a large influence on how confidence was manipulated. A mixed-effects repeated measures ANOVA, with within-subjects factor time, and between-subjects factors neurofeedback, and order, clarifies this finding, as it resulted in a significant interaction between the three factors ($F_{1,16}$ = 4.769, $P$ = 0.044). Furthermore, the factor time ($F_{1,16}$ = 4.623, $P$ = 0.047) and the interaction between time and neurofeedback ($F_{1,16}$ = 18.050, $P$ = 0.001) both had significant effect on the dependent variable, confidence.

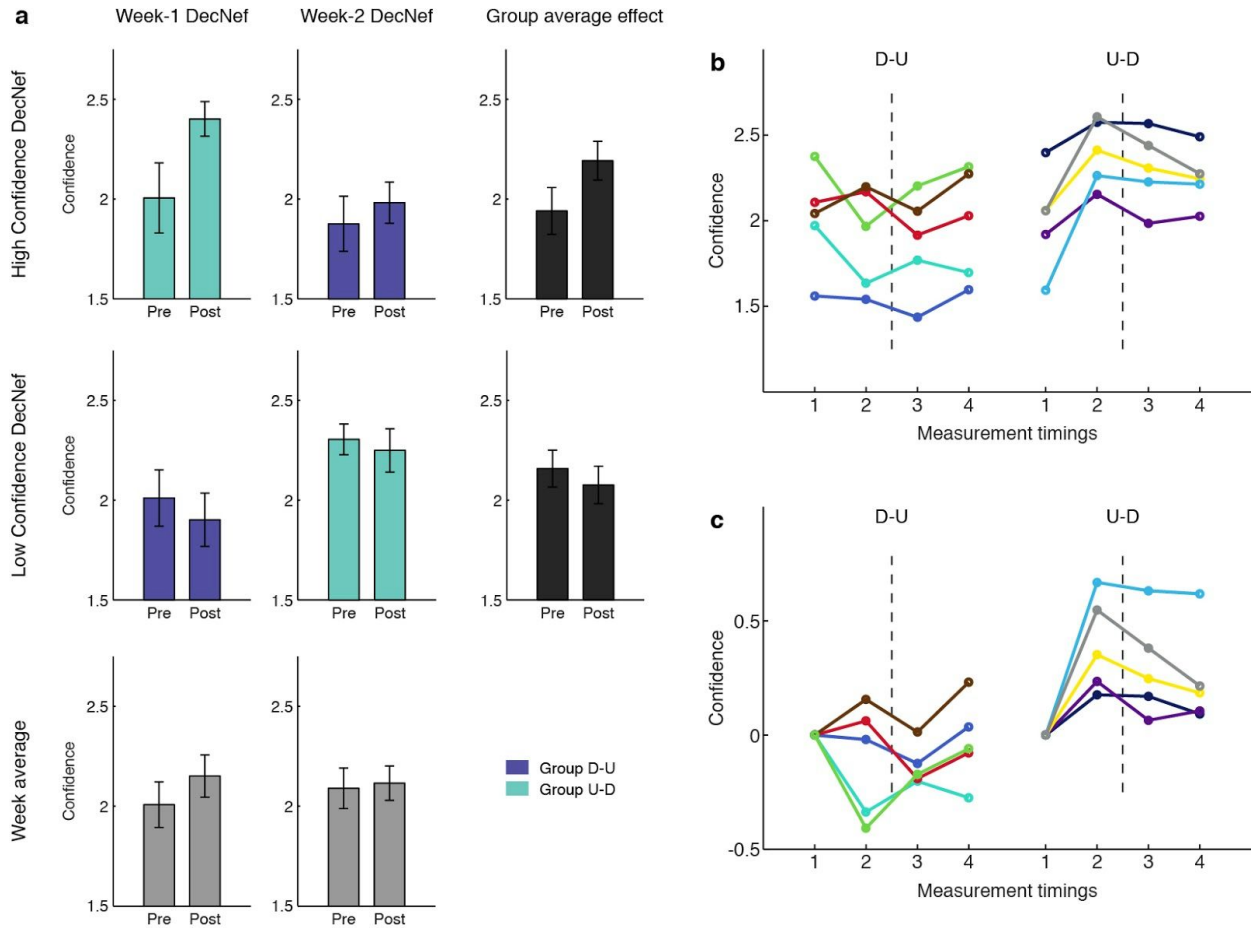

**Figure 2.** Confidence ratings in psychophysical Pre- and Post-Tests. (a) Each group induced both High and Low Confidence, in two separate sessions: group D-U with the order Low- then High-Confidence (Down then Up), and group U-D High- then Low Confidence. Results are color-coded: purple for single group results of group D-U and green for group U-D. Average across groups, representing mere order effect, is depicted in gray, at the bottom, while the grand average per condition (High and Low Confidence), is in black, on the right side. It is apparent that the confidence changes in the first week strongly influence the confidence level in the second week. The Pre- level in the second week remains indeed very close to the previous Post- level attained in the first week. Furthermore, neurofeedback in the first week seems to have a stronger impact than in the second week, and this holds when comparing both High with Low Confidence DecNef. For each bar, N=5. Error bars represent s.e.m. (b-c) Subjects' individual confidence data with the four measurement timings, organized in groups (D-U and U-D), both raw (b) and 0-aligned (c) confidence. Timings 1 and 2 correspond to Pre- and Post-Test in the first week, respectively; while timings 3 and 4 to Pre- and Post-Test in the second week DecNef. The dotted line represents the week-long interval between the two sessions (DecNef in week 1 and DecNef in week 2). Each colored line represents day-averaged data from one subject (total, N=10).

---

The results at the group level (Fig. 2A) were mirrored at the individual level (Fig. 2B, C). Fig. 2B shows that the initial value varied across subjects, but also that on average the initial point was the same for both groups and that changes had a clear common trend across subjects. Thus, in Fig. 2C, data were realigned to the same starting point, centered on zero. In more detail, Fig. 2C suggests that 7/10 cases in the Down-DecNef and 9/10 cases in the Up-DecNef showed changes in confidence in the expected directions. The Phi coefficient (a measure of the correlation between two vectors of binary variables) computed between real and expected directional changes was strongly significant ($\phi$ = 0.612, $P$ = 0.0041). That is, confidence increased in Up-DecNef weeks, and decreased in Down-DecNef weeks.

A concept that has been extensively studied in motor learning and, to a lesser extent, in perceptual learning, is learning interference. In a classic motor learning interference paradigm, Krakauer *et al.* (Krakauer et al. 1999) showed that learning of another kinematic or dynamic model with conflicting sensorimotor mappings interfered with the consolidation of previously learned models of the same type. Similarly, in this study, we propose that DecNef training also induced an anterograde interference effect, where learning of task B was partly prevented by the previous learning of task A. This effect will be more rigorously examined later.

To effectively analyze the changes in confidence, the two DecNef sessions for the two groups, D-U1, D-U2, and U-D1, U-D2, respectively, are presented as deltas (Fig. 3A). As expected, average changes were positive for Up and negative for Down-DecNef: U-D1 data was significantly different from zero (one-tail t-test, $t_{(4)}$ = 4.253, $P$ = 0.0067, uncorrected; $P$ = 0.026, corrected for multiple comparisons), as well as D-U2 (one-tail t-test, $t_{(4)}$ = 2.188, $P$ = 0.0469, uncorrected, $P$ = 0.14, corrected). Both D-U1 (one-tail t-test, $t_{(4)}$ = -0.978, $P$ = 0.192, uncorrected, $P$ = 0.192, corrected) and U-D2 (one-tail t-test, $t_{(4)}$ = -1.626, $P$ = 0.0896, uncorrected, $P$ = 0.179, corrected) were not statistically different from zero. The contrast between D-U1 and U-D1 yielded a statistically significant difference (one-tail t-test, $t_{(4)}$ = -3.822, $P$ = 0.0094, uncorrected, $P$ = 0.0468, corrected). This result is of great importance, because in the two instances only the neurofeedback sign was different, while all other behavioral schemes were the same, and yet different results were obtained. Thus, DecNef purely induced bidirectional confidence changes, and these confidence changes were not caused by general effects of monetary rewards or repeated exposure to random dot stimuli, which are common experimental components of both up and down DecNef. Furthermore, mean differences between U-D1 and U-D2 (one-tail t-test, $t_{(4)}$ = 4.228, $P$ = 0.0067, uncorrected, $P$ = 0.0402, corrected), D-U2 and U-D1 (one-tail t-test, $t_{(4)}$ = -3.661, $P$ = 0.0108, uncorrected, $P$ = 0.0431, corrected) yielded statistically significant results before and after multiple comparisons correction. It should be noted that, although they did not survive a multiple comparisons correction, even the differences between D-U1 and D-U2 (one-tail t-test, $t_{(4)}$ = -2.512, $P$ = 0.033, uncorrected, $P$ = 0.0989, corrected), and D-U2 and U-D2 (one-tail t-test, $t_{(4)}$ = 2.296, $P$ = 0.0416, uncorrected, $P$ = 0.0833, corrected) were initially significant. Since one could argue that the conditions between each comparisons are different, because they entail different DecNef directions and therefore assumptions, it is noteworthy to see that most of the confidence differences were significantly distinct.

Fig. 3B plots the ratios of the above differences U-D2/D-U1, D-U2/U-D1 and the average of the two. Because these values are positive and less than 1, the second week effect was smaller than the first week effect, an outcome that is likely due to anterograde learning interference. The first two values being relatively similar, the interference effect did not seem to be dependent upon Up-Down or Down-Up sequence.

Therefore, considering that differences in the first and second week of DecNef are quantifiable and can be ascribed to a specific hypothesis, pure Up and Down effects can be computed by applying a simple correction. Reduced Up and Down-DecNef effects in the second week can be corrected for the first week effect by dividing them by the average ratio (Fig. 3C). Both Up and Down effects were statistically significantly different from zero (one-tail t-test, Up-DecNef $t_{(9)}$ = 4.390, $P$ = 0.0009, Down-DecNef $t_{(9)}$ = -1.876, $P$ = 0.0467), and they were different from each other (one-tail t-test, $t_{(9)}$ = 4.315, $P$ = 0.001), suggesting a specific DecNef effect to neurofeedback signs.

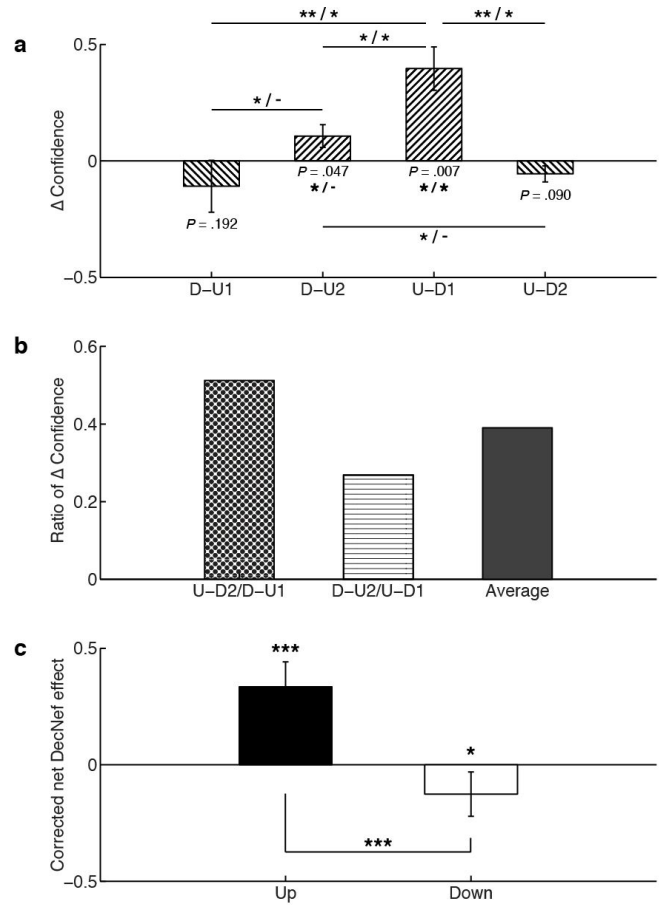

**Figure 3.** Summary statistics for confidence changes. (a) Δ confidence is given by the difference [Post-Test - Pre-Test] of confidence average. D-U1 is the confidence difference in group D-U, DecNef session 1, D-U2 the confidence distance in group D-U, DecNef session 2, U-D1 group U-D, DecNef session 1, and U-D2 group U-D, DecNef session 2. Asterisks to the left of the slashes represent uncorrected significant differences, while asterisks to the right, significant differences after correction for multiple comparisons ( * $P$ < 0.05, ** $P$ < 0.01, *** $P$ < 0.005). (b) Ratios of Δ confidence measured as confidence change in the second week divided by the confidence change in the first week, for both Up- and Down-DecNef. The right column indicates the average of the two ratios, used in the analysis displayed in the next panel. (c) Corrected net DecNef effect: the correction is computed by dividing the second week effect by the averaged ratios of Δ confidence. This term accounts for the anterograde learning

interference, decreasing the DecNef effect in the second week. Both Up- and Down-DecNef are thus significantly different from zero.

From a mathematical standpoint, in order to best capture the different components of DecNef effects while accounting for both Up- and Down neurofeedback, we fitted a system of nonlinear parametric equations with four global parameters (explained in greater detail in the mathematical modeling part of the Materials and methods section). The four global parameters were selected in light of the summary statistics results displayed in Fig. 3. Global parameters hence were the initial (first week) absolute change in confidence by Up-DecNef effect ($\Delta$), the ratio of the weaker Down-DecNef effect compared with that of stronger Up-DecNef effect ($\varepsilon$), anterograde learning interference ($ɣ$), and the memory loss effects between DecNef sessions ($α$). Importantly, we fitted various alternative models, where some of the parameters had fixed values (such as = 1 or = 0) to account for full effects, or the lack of effects, in order to compare and infer which aspects of DecNef were likely to play a significant role in determining the resulting confidence changes. Simpler models, that did not assume directionality in confidence changes, or other interpretations, included a constant-confidence model, constant-within-week model, and first-grade polynomial models. In order to compare various models, we used the corrected (second-order) Akaike Information Criterion (AICc), which allocates more importance to the principle of parsimony, and gives higher penalty to more complex models (i.e., with larger number of parameters). Indeed, since the dataset is finite, and the ratio k/n (number of parameters / number of samples) > 1/40, AICc is strongly recommended to avoid model selection bias (Burnham & Anderson 2002)).

When using AIC (and, by extension, AICc), the best model has the most negative (or lowest) score. Absolute values of AICs are not really informative *per se*, and to effectively compare models we use the distance from the best model ($\Delta_{AICc}$), and the likelihood of each model being the best model, computed as Akaike weights. Normally, a $\Delta^i_{AICc} < 2$ means the *i*th model cannot be ruled out and has a conspicuous likelihood of being the best model with a different dataset. Accordingly, we computed AICc scores, $\Delta_{AICc}$, and $w_i$ (reported in table 1). As shown in the table, there are three models that can be considered essentially as good as the best model, since the distance between two of them and the most negative AICc is < 2. Furthermore, three more models had $\Delta_{AICc}$ < 6, indicating these had a very low likelihood, but nevertheless marginal validity. All other models had distances greater than 10 from the best model, and importantly, among them simpler models such as constant-confidence, within-week constant confidence, and first-grade polynomial models all performed very poorly in fitting the data. These models with $\Delta_{AICc}$ > 10 are sufficiently poorer than the best AIC model as to be considered implausible (Burnham & Anderson 2002).

Since model selection uncertainty exists, with very similar AICc values ($\Delta_{AICc}$ < 2 compared to the most negative AICc), a formal solution is to apply model-averaging, where each parameter present in the selected models is estimated according to a weighted average based on their corresponding Akaike weights. For model averaging, we used all models for which $\Delta_{AICc}$ < 2, keeping the estimated parameters in the same initial scale and focusing the averaging process on the subset of highly likely models.

TABLE 1. AICc analysis

| Model (est. parms) | Fixed parms | AICc | $\Delta_i$ | $w_i$ |
|---|---|---|---|---|
| *Const. confidence* | k | -70.6447 | 24.3397 | 0 |
| *Polynomial* ($\alpha_1, \alpha_2$) | k | -67.9402 | 27.0442 | 0 |
| *Polynomial* ($\alpha_1, \beta_1, \alpha_2, \beta_2$) | $k_1$ $k_2$ | -75.3869 | 19.5975 | 0 |
| *Sub – model* ($\Delta$) | $\alpha=1$ $\epsilon=0$ $\gamma=1$ | -81.6244 | 13.3600 | 0.0005 |
| *Sub – model* ($\Delta, \epsilon$) | $\alpha=1$ $\gamma=1$ | -91.5062 | 3.4782 | 0.0637 |
| *Sub – model* ($\Delta, \epsilon, \alpha$) | $\gamma=1$ | -89.0275 | 5.9569 | 0.0185 |
| *Sub – model* ($\Delta, \epsilon, \gamma$) | $\alpha=0$ | -77.3589 | 17.6255 | 0.0001 |
| *Sub – model* ($\Delta, \gamma, \alpha$) | $\epsilon=0$ | -91.4285 | 3.5559 | 0.0613 |
| **Sub – model** ($\mathbf{\Delta, \epsilon, \gamma}$) | $\alpha=1$ | -94.9402 | 0.0442 | 0.3549 |
| **Full – model** ($\mathbf{\Delta, \epsilon, \gamma, \alpha}$) | | -93.0529 | 1.9315 | 0.1381 |
| **Sub – model** ($\mathbf{\Delta, \epsilon, \alpha}$) | $\gamma=0$ | -94.9844 | 0 | 0.3629 |

Fig. 4A reports individual data, as well as fits of the three best models ($\Delta_{AICc}$ < 2, full model with 4 global parameters - α, ε, ɣ, Δ; submodel with ɣ=0, submodel with α=1), and the simplest model, where confidence does not change and is akin to a 1-k model, with the only parameter being the confidence group average. As can be deduced from the figure, each model provides a good fit, and therefore a high likelihood of being the best one at describing the empirical data. Conversely, a simpler model, where confidence is assumed to be constant, provides a very poor fit to the data, showing that DecNef was indeed successful in inducing confidence changes, and that the no-change model is implausible.

Model-averaged estimates of the 4 parameters suggest that all of them are different from zero (Fig. 4B), and thus impact the effects of DecNef.

The delta parameter was 0.37; thus Up-DecNef on the first week increased confidence by 0.37 (~20% change in confidence). Alpha was 0.83, hence on average only 17% of the first week effect was lost during the one week interval due to memory decay. Epsilon was -0.35, thus the Down-DecNef effect was opposite in its sign and 35% of the magnitude of that of the Up-DecNef. Gamma was 0.19, and thus, due to anterograde learning interference, the second week effect was only 19% of that of the first week. Our nonlinear modeling robustly indicates that there exist both Up- and Down-DecNef effects, anterograde learning interference, and preservation of memory between sessions one-week apart.

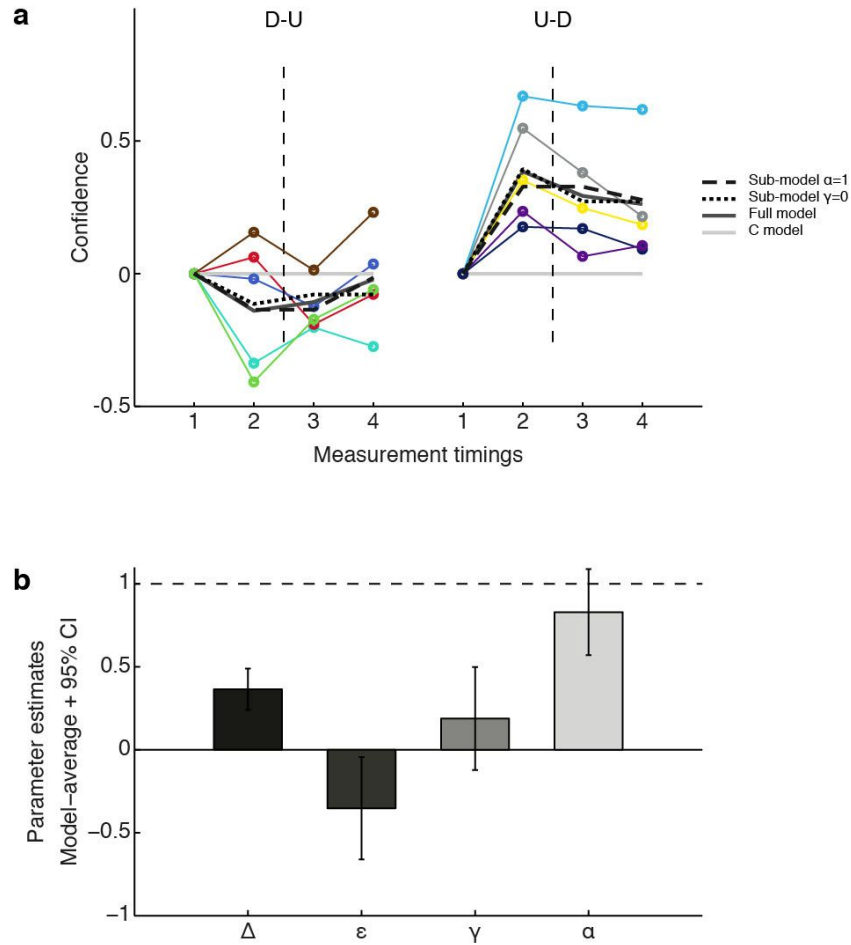

**Figure 4.** Confidence changes modeled with global parameters and nonlinear functions. (a) Individual and group-modeled data are presented, 0-aligned, with the same color-codes and presentation rules as in fig. 3. Thick lines represent different model fits, darker colors being better fits. The light gray line is the simplest model, assuming constant confidence (1 parameter model). The three other fits are, respectively, the best model, for which the computed AICc was the most negative, and models having a $\Delta_{AICc} < 2$ from the best model. Actual data from each participant is shown as a thin colored line. (b) Global parameters estimates resulting from model averaging. For each model, based on the $\Delta_{AICc}$, Akaike weights were computed (the likelihood of the *i*th model being the best model). Each parameter was then evaluated as the weighted average of single models estimates. Error bars represent the 90% confidence intervals, computed from the unconditional standard error.

In the second part of the modeling analysis, we considered all data points, and model population's individual fits with nonlinear equations with *global* and *local* parameters. This approach is warranted by the results from the global model-averaging, where all four parameters have high likelihood of being different from zero. This aspect, albeit not entirely backed by the confidence intervals (CI for $\gamma$ includes 0), is nevertheless supported by the internal variance of the models: these variances are very low, indicating high stability within a model's parameter space. Furthermore, if the models are robust and explain the same phenomenons in DecNef, parameter estimates should converge to similar solutions.

In the second model, global parameters were the weaker Down-DecNef effect ($\varepsilon$), anterograde learning interference ($\gamma$), and the memory loss effects between DecNef sessions ($\alpha$), and local parameters were the individual initial absolute change in confidence for Up-DecNef effect ($\Delta_i$), and the individual initial point of confidence ($b_i$). Fig. 5A shows the 23 parameter model fit to the raw data, and provides further support for Fig. 4 conclusions. Parameter estimate values (Fig. 5B) are in good agreement in both modeling instances (Fig. 4B and 5B), underscoring the presence and generality of the effects they represent, as well as supporting the robustness of the model analyses.

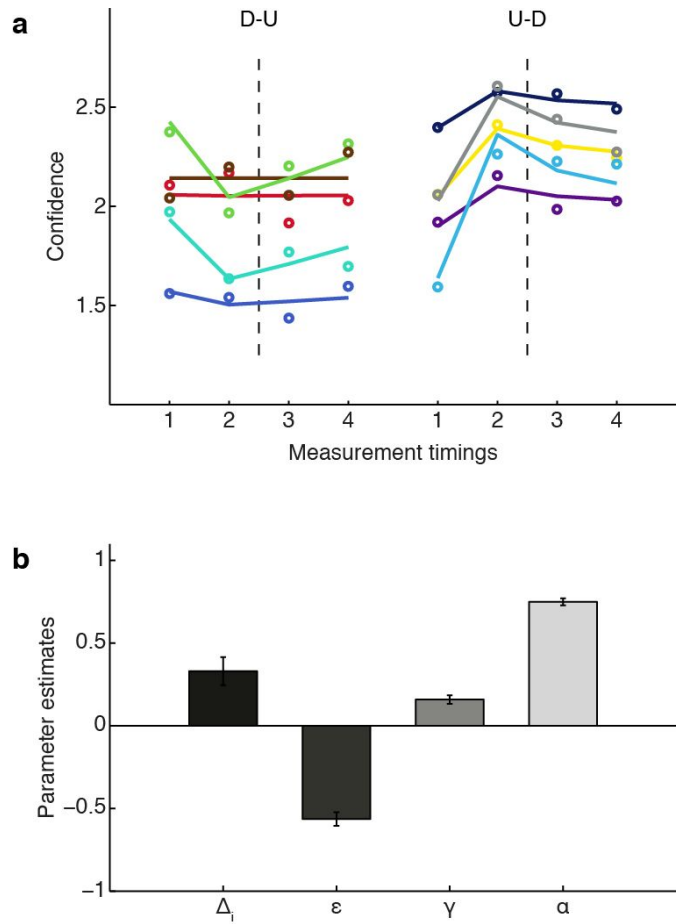

**Figure 5.** Global and local parameters modeling: the model is based on the parameters obtained from the previous model-averaging step using global models. This full model accounts for all four major effects of DecNef, as exemplified in the parameter estimates plot in the panel below: $\Delta_i$, individual differences in Up-DecNef confidence change; $\varepsilon$, reduced Down-DecNef effect; $\gamma$, anterograde learning interference resulting in reduced second week DecNef effect; $\alpha$, learning persistence (memory loss in the week-long interval). (a) Thick lines are individual fits, while dots are the actual measured data. Color-codes and presentation rules are the same as in Figs 3 and 5. (b) Model estimates. Error bars are standard deviation for $\Delta_i$, and CI for $\varepsilon$, $\gamma$, and $\alpha$. D-U: Down-Up DecNef order (Low- then High Confidence DecNef), U-D: Up-Down DecNef order.

To conclude, by utilising the gamma parameter, representing anterograde learning interference, computed in the global modeling approach through model averaging, we show that DecNef was effective in both directions (Fig. 6). That is, simply discounting the weaker second week effect by correcting (dividing) the individual pre- post- confidence change from the second week by the gamma parameter indeed leads to a significance level that is statistically relevant for both increase and decrease of confidence.

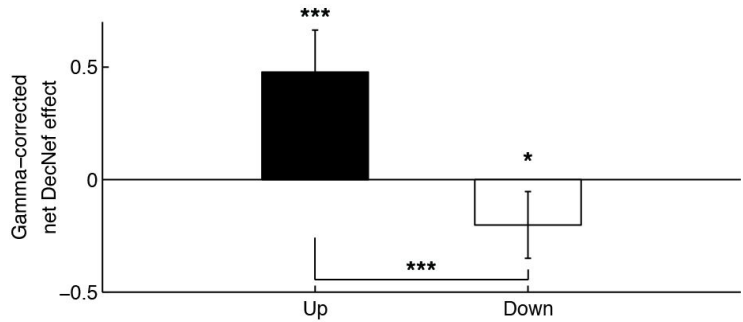

Figure 6. Pure effects of Up- and Down-DecNef trainings. The second week confidence difference between Pre- and Post-tests were individually corrected by dividing the difference by the gamma parameter computed in the global model approach by model averaging ($\gamma = 0.19$). One-tail t-test for Up-DecNef ($t_{(9)}$ = 3.64, $P$ = 0.0027), Down-DecNef ($t_{(9)}$ = -1.92, $P$ = 0.0433), as well as direct comparison of the two pure effects ($t_{(9)}$ = 3.47, $P$ = 0.0034), were all significant, evidencing the bidirectional effectiveness of DecNef.

## Discussion

We hypothesized that DecNef could be successfully used to induce changes in a behavioral trait (confidence) for two opposing directions and within the same subjects. Furthermore, if DecNef were based on a learning process in the behavioral dimension probed, we expected an anterograde learning interference. To directly examine these hypotheses, we constructed individual decoders based on confidence judgements in a visual perceptual task, and participants then induced such multi-voxel activation patterns in two DecNef sessions, i.e., one for High Confidence and one for Low Confidence induction. Importantly, each session was separated at least one week apart, in order to best capture possible learning-maintenance effects.

Our modeling results support the assumptions that confidence changes due to DecNef followed a specific pattern, where (1) DecNef was successful in both opposing directions, but that Up-DecNef effect was more pronounced than Down-DecNef, (2) the acquired confidence level at the end of the first week was subjected to only small degradation effect, (3) there existed a strong anterograde interference onto the next DecNef session by the first week DecNef. The consequence was a stronger effect of neurofeedback in the first week as compared to the second week.

The anterograde interference in learning that emerges when training participants to induce activation patterns that correspond to different (opposite) behavioral variables sequentially is important. This effect is remarkable, because it implies that any manipulation through DecNef is likely relying on long term changes akin to sensory motor learning or perceptual learning, thus providing an additional support to previous empirical finding in rt-fMRI (Shibata et al. 2011; Megumi et al. 2015). Specifically, these behavioral changes may be deeply ingrained due to the instrumental conditioning that is subtending such learning processes.

In motor and visuomotor learning, a conspicuous literature has explored the effect of opposing tasks on the dynamics and modalities of the learning processes (Brashers-Krug et al. 1996; Krakauer et al. 1999; Tong et al. 2002; Osu et al. 2004). Both retrograde and anterograde mechanisms have been suggested to mediate interference in visuomotor learning (Tong et al. 2002; Miall et al. 2004; Krakauer et al. 2005). The vast majority of previous studies have addressed interference in learning within short delays (typically, between a few minutes and up to 24h-48h), but, nevertheless, several studies have reported interference effects even after 1 week (Caithness et al. 2004; Krakauer et al. 2005), thus corroborating our results. Furthermore, anterograde interference is thought to have substantially larger effects as compared to retrograde interference (Sing & Smith 2010). It is therefore not surprising that the interference found in this study resulted in a second week effect of only ~20% of the size of the first week, a significant decrease.

To conclude, we established the causal nature of DecNef, as bidirectional brain manipulation led to bidirectional behavioral change, in that DecNef was effective in both increasing and decreasing confidence. By finding a strong anterograde interference in the behavioral results, DecNef can be seen as a natural learning, likely close to sensory motor learning and perceptual learning. From a translational viewpoint, DecNef is effective since only two days training were sufficient to maintain 80% of the learned level after one week elapsed. Once acquired, cancelling out DecNef effects was difficult (only 20%), possibly requiring more days of induction. Up and down modulations are asymmetrical, suggesting some differential basic neural mechanisms for confidence encoding.

# References

Akaike, H., 1974. A new look at the statistical model identification. *IEEE transactions on automatic control*, 19(6), pp.716–723.

Amano, K., Shibata, K., Kawato, M., Sasaki, Y., Watanabe, T., 2016. Creation of no-aftereffect-based associative learning of color and orientation without presenting color by decoded fMRI neurofeedback. *Journal of Vision*, in press.

Birbaumer, N., Ruiz, S. & Sitaram, R., 2013. Learned regulation of brain metabolism. *Trends in cognitive sciences*, 17(6), pp.295–302.

Brainard, D.H., 1997. The Psychophysics Toolbox. *Spatial vision*, 10(4), pp.433–436.

Brashers-Krug, T., Shadmehr, R. & Bizzi, E., 1996. Consolidation in human motor memory. *Nature*, 382(6588), pp.252–255.

Bray, S., Shimojo, S. & O'Doherty, J.P., 2007. Direct Instrumental Conditioning of Neural Activity Using Functional Magnetic Resonance Imaging-Derived Reward Feedback. *Journal of Neuroscience*, 27(28), pp.7498–7507.

Burnham, K.P. & Anderson, D.R., 2002. *Model Selection and Multimodel Inference: A Practical Information-Theoretic Approach*, Springer Science & Business Media.

Caithness, G. et al., 2004. Failure to consolidate the consolidation theory of learning for sensorimotor adaptation tasks. *The Journal of neuroscience: the official journal of the Society for Neuroscience*, 24(40), pp.8662–8671.

deBettencourt, M.T. et al., 2015. Closed-loop training of attention with real-time brain imaging. *Nature neuroscience*, 18(3), pp.470–475.

deCharms, R.C., 2008. Applications of real-time fMRI. *Nature reviews. Neuroscience*, 9(9), pp.720–729.

deCharms, R.C. et al., 2004. Learned regulation of spatially localized brain activation using real-time fMRI. *NeuroImage*, 21(1), pp.436–443.

Fleming, S.M. et al., 2010. Relating introspective accuracy to individual differences in brain structure. *Science*, 329(5998), pp.1541–1543.

Kamitani, Y. & Tong, F., 2005. Decoding the visual and subjective contents of the human brain. *Nature neuroscience*, 8(5), pp.679–685.

Kiani, R. & Shadlen, M.N., 2009. Representation of confidence associated with a decision by neurons in the parietal cortex. *Science*, 324(5928), pp.759–764.

Koralek, A.C. et al., 2012. Corticostriatal plasticity is necessary for learning intentional neuroprosthetic skills. *Nature*, 483(7389), pp.331–335.

Krakauer, J.W., Ghez, C. & Ghilardi, M.F., 2005. Adaptation to visuomotor transformations: consolidation, interference, and forgetting. *The Journal of neuroscience: the official journal of the Society for Neuroscience*, 25(2), pp.473–478.

Krakauer, J.W., Ghilardi, M.F. & Ghez, C., 1999. Independent learning of internal models for kinematic and dynamic control of reaching. *Nature neuroscience*, 2(11), pp.1026–1031.

Megumi, F. et al., 2015. Functional MRI neurofeedback training on connectivity between two regions induces long-lasting changes in intrinsic functional network. *Frontiers in human neuroscience*, 9.

Miall, R.C., Jenkinson, N. & Kulkarni, K., 2004. Adaptation to rotated visual feedback: a re-examination of motor interference. *Experimental brain research.* 154(2), pp.201–210.

Osu, R. et al., 2004. Random presentation enables subjects to adapt to two opposing forces on the hand. *Nature neuroscience*, 7(2), pp.111–112.

Rounis, E. et al., 2010. Theta-burst transcranial magnetic stimulation to the prefrontal cortex impairs metacognitive visual awareness. *Cognitive neuroscience*, 1(3), pp.165–175.

Scheinost, D. et al., 2013. Orbitofrontal cortex neurofeedback produces lasting changes in contamination anxiety and resting-state connectivity. *Translational psychiatry*, 3, p.e250.

Seitz, A.R. et al., 2005. Task-specific disruption of perceptual learning. *Proceedings of the National Academy of Sciences of the United States of America*, 102(41), pp.14895–14900.

Shadlen, M.N. & Newsome, W.T., 2001. Neural basis of a perceptual decision in the parietal cortex (area LIP) of the rhesus monkey. *Journal of neurophysiology*, 86(4), pp.1916–1936.

Shibata, K. et al., 2016. Differential activation patterns in the same brain region led to opposite emotional states. Available at: http://arxiv.org/abs/1603.01351.

Shibata, K. et al., 2011. Perceptual learning incepted by decoded fMRI neurofeedback without
31

stimulus presentation. *Science*, 334(6061), pp.1413–1415.

Sing, G.C. & Smith, M.A., 2010. Reduction in learning rates associated with anterograde interference results from interactions between different timescales in motor adaptation. *PLoS computational biology*, 6(8).

Sulzer, J. et al., 2013. Real-time fMRI neurofeedback: progress and challenges. *NeuroImage*, 76, pp.386–399.

Tong, C., Wolpert, D.M. & Flanagan, J.R., 2002. Kinematics and dynamics are not represented independently in motor working memory: evidence from an interference study. *The Journal of neuroscience: the official journal of the Society for Neuroscience*, 22(3), pp.1108–1113.

Veit, R. et al., 2012. Using real-time fMRI to learn voluntary regulation of the anterior insula in the presence of threat-related stimuli. *Social cognitive and affective neuroscience*, 7(6), pp.623–634.

Weiskopf, N. et al., 2003. Physiological self-regulation of regional brain activity using real-time functional magnetic resonance imaging (fMRI): methodology and exemplary data. *NeuroImage*, 19(3), pp.577–586.

Weiskopf, N. et al., 2004. Principles of a brain-computer interface (BCI) based on real-time functional magnetic resonance imaging (fMRI). *IEEE transactions on biomedical engineering*, 51(6), pp.966–970.

Yamashita, O. et al., 2008. Sparse estimation automatically selects voxels relevant for the decoding of fMRI activity patterns. *NeuroImage*, 42(4), pp.1414–1429.

Yotsumoto, Y. et al., 2009. Interference and feature specificity in visual perceptual learning. *Vision research*, 49(21), pp.2611–2623.